\documentclass[12pt]{article}
\usepackage{amsmath}
\newlength{\abstractwidth}
\begin{document}
\thispagestyle{empty}
\pagestyle{plain}
\def\beq{\begin{eqnarray}}
\def\eeq{\end{eqnarray}}
\def\nn{\nonumber\\}
\def\heq{\,\hat=\,}
\newcommand{\const}{\mbox{const}}
\renewcommand{\thefootnote}{\fnsymbol{footnote}}
\renewcommand{\thanks}[1]{\footnote{#1}} 
\newcommand{\starttext}{
\setcounter{footnote}{0}
\renewcommand{\thefootnote}{\arabic{footnote}}}
\begin{titlepage}
\bigskip
\hskip 3.7in\vbox{\baselineskip11pt
\hbox{CGPG-01/1-1}}
\bigskip\bigskip\bigskip\bigskip

\begin{center}
{\large \bf Black hole entropy calculations based on symmetries
\\[.4cm]}
\end{center}

\bigskip\bigskip
\bigskip\bigskip

\centerline{\bf Olaf Dreyer\footnote{dreyer@phys.psu.edu}, Amit
Ghosh\footnote{ghosh@gravity.phys.psu.edu} and Jacek
Wi\'sniewski\footnote{jacek@phys.psu.edu}}
\medskip
\begin{center}
Center for Gravitational Physics and Geometry, Department of
Physics,\\ The Pennsylvania State University,
University Park, PA 16802-6300, USA\\
\end{center}

\bigskip

\abstract{Symmetry based approaches to the black hole entropy
problem have a number of attractive features; in particular they are
very general and do not depend on the details of the quantization
method. However we point out that, of the two available approaches,
one faces conceptual problems (also emphasized by others), while the
second contains certain technical flaws. We correct these errors and,
within the new, improved scheme, calculate the entropy of
3-dimensional black holes.  We find that, while the new symmetry
vector fields are well-defined on the ``stretched horizon,'' and lead
to well-defined Hamiltonians satisfying the expected Lie algebra, they
fail to admit a well-defined limit to the horizon. This suggests that,
although the formal calculation can be carried out at the classical
level, its real, conceptual origin probably lies in the quantum
theory.}



\vspace{1cm}

\begin{flushleft}
\today
\end{flushleft}
\end{titlepage}
\starttext
\setcounter{footnote}{0}

\section{Introduction}

A microscopic derivation of black hole entropy has been one of the
greatest theoretical challenges for any candidate quantum theory of
gravity. String theory in case of some extremal and {\em near}-extremal
black holes \cite{Strom} and canonical quantum gravity in case of
general non-rotating black holes \cite{Abhay} have produced
very interesting results in this direction.

As an alternative to both these approaches, a set of very
attractive ideas was suggested by Strominger and Carlip
\cite{Strom2, Carlip1, Carlip2} over the last few years.
Motivated by some earlier works \cite{BH, BH1, Regge} on the relation
between symmetries and Hamiltonians, these authors argued that
states of a quantum black hole should belong to a multiplet of a
representation of a suitable Lie algebra. Counting the number of
states in the multiplet would then provide the black hole entropy.
The Virasoro algebra has been proposed as a natural candidate for
symmetries in this context.

An attractive feature of these alternative approaches is that they are
not tied to the details of any specific model of quantum gravity. Even
more strikingly, the central objects in this construction, namely the
Virasoro algebra, central charge etc.  appear already at the {\em
classical} level through the Poisson bracket algebra. The Planck
length in the expression of the entropy arises only from replacing
Poisson brackets by appropriate quantum commutators. Being essentially
classical, the scheme is quite robust and in principle applicable to
black holes in any space-time dimension.

The first work \cite{Strom2} applies this idea to 2+1 dimensions in
the context of the BTZ black hole \cite{BTZ}. The symmetries, however,
are taken from a previous analysis \cite{BH1} which is tailored to
\textit{asymptotic infinity} rather than black hole
horizon. Therefore, it is not apparent why these symmetries are
relevant for the black hole in the space-time interior. For example,
in asymptotically flat, 4-dimensional space-times, the symmetry group
at (null) infinity is always the Bondi-Metzner-Sachs group,
irrespective of the interior structure of the space-time. Thus, the
results of \cite{Strom2} are equally applicable to a star that has
similar asymptotic behavior as that of the black hole. Subsequently,
Carlip improved on this idea significantly by making the symmetry
analysis in the near-horizon region. Conceptually this approach is
much more satisfactory in that the black hole geometry is now at the
forefront. However, at the technical level, this work \cite{Carlip2}
appears to have some important limitations.  The purpose of the
present paper is to elucidate and discuss these technical problems in
some detail, and then to present a consistent calculation which
correctly implements the general ideas of \cite{Strom2, Carlip2} by
a careful treatment of all the relevant technical issues.

The organization of the paper is as follows. In section 2 we
discuss the technical framework set up in \cite{Carlip2} and point
to the difficulties that arise in the implementation of the ideas
mentioned above. In section 3 we investigate two different sets of
symmetries. In 3.1 we consider symmetries that are defined
intrinsically on the horizon and see if a central charge can be
obtained. (To find these symmetries, as suggested in
\cite{Carlip2}, we use the isolated horizon framework
\cite{letter}.) We find that the answer is in the negative. In
3.2, we then consider potential symmetry vector fields defined in
a neighborhood of the horizon as in \cite{Carlip2}. In 3.3, we
find the corresponding Hamiltonians, and calculate the
corresponding Poisson brackets. From these we read off the central
charge. We conclude in 3.4 with a calculation of the entropy. The
discussion of the last three sub-sections can be regarded as a
careful reworking of the ideas introduced by Strominger and
Carlip. We find that the entropy is indeed proportional to the
area but the proportionality factor differs from the one of
Bekenstein and Hawking by a factor of $\sqrt{2}$. Perhaps more
importantly, although the Poisson brackets between the
Hamiltonians are well-defined, the symmetry vector fields
underlying this calculation fail to admit a well-defined limit to
the horizon. These issues are discussed in section 4. Some
technical details relevant to section 3.1 are given in the
appendix.

For concreteness, we work in 2+1 dimensions. However, the framework
should admit a straightforward generalization to arbitrary space-time
dimensions.

\section{Re-examination of the symmetry based calculations in 3 dimensions}

In the standard conventions (with $8G =1$), the line-element of BTZ
black hole in the {\em Eddington-Finkelstein} coordinates is given by
\beq
&&ds^2=-N^2dv^2+2dvdr+r^2(d\phi+N^\phi dv)^2\;,\nn
&&N^2=-M+\frac{r^2}{\ell^2}+{\frac{J^2}{4r^2}}\;,\;N^\phi=-\frac{J}
{2r^2}\;.\label{metric}
\eeq
Here, $J$ and $M$ are two real parameters and $\ell$ is related to
the negative cosmological constant as $\Lambda\ell^2=-1$. The
black hole has a Killing-horizon at $r=r_{+}$ defined by
$N^2(r_+)=0$, or \beq r_+^2= \frac{1}{2}
M\ell^2\bigg[1+\sqrt{1-\frac{J^2}{M^2\ell^2}}\bigg]^{1/2},\quad
|J| \leq M\ell\;. \eeq For the purpose of calculations it is
convenient to introduce Newman-Penrose like basis in 2+1
dimensions which has two null vectors $l^a$ and $n^a$ and a
space-like vector $m^a$ (all real). They satisfy the relations
\beq l\cdot l=n\cdot n=l\cdot m=n\cdot m=0\;,\;-l\cdot n=m\cdot
m=1\;.\label{basis} \eeq The 2+1 dimensional metric can be
expressed in such a basis as $g_{ab}=-2l_{(a}n_{b)}+m_am_b$. The
corresponding inverse metric is $g^{ab}=-2l^{(a}n^{b)}+m^am^b$. In
the rest of the paper we will assume that we have chosen the triad
$l, n$, and $m$ in such a way that the vectors $l$ and $m$ are
tangent to the horizon at the horizon. For the metric
(\ref{metric}) a convenient choice of the basis vector fields is
\beq
l=\partial_v+\frac{1}{2}N^2\partial_r-N^\phi\partial_\phi\;,\;
n=-\partial_r\;,\;m=\frac{1}{r}\partial_\phi\;.\label{vectors}
\eeq and the corresponding one-forms that span the dual-basis are
\beq l=-\frac{1}{2}N^2dv+dr\;,\;n=-dv\;,\;m=rN^\phi dv+rd\phi\;.
\label{1form}\eeq The covariant derivatives of the one-forms, like
$\nabla_al_b$, can be expressed solely in terms of the one-forms
and the so-called Newman-Penrose coefficients (See e.g.
\cite{Stewart}; an exposition of the formalism in 2+1 dimensions
can be found in the appendix of \cite{ihtpo})

\begin{equation}\nonumber
\nabla_al_b=-\epsilon n_al_b+\tilde\kappa n_am_b-\gamma l_al_b
+\tau l_am_b+\alpha m_al_b-\rho m_am_b
\end{equation}
\begin{equation}\label{diffid2}
\nabla_an_b=\epsilon n_an_b-\pi n_am_b+\gamma l_an_b-\nu l_am_b
-\alpha m_an_b+\mu m_am_b
\end{equation}
\begin{equation}\nonumber
\nabla_am_b=\tilde\kappa n_an_b-\pi n_al_b+\tau l_an_b -\nu
l_al_b-\rho m_an_b+\mu m_al_b
\end{equation}

where, for the metric (\ref{metric}) and the tetrad
(\ref{vectors}) the coefficients are given by
\begin{equation}\nonumber
  \epsilon={r\over\ell^2}-r(N^\phi)^2,\ \rho = -{1\over 2r}N^2,\ %
  \mu=-{1\over r}
\end{equation}
\begin{equation}\label{coeff}
    \alpha\ =\ \tau\ =\ \pi = N^\phi
\end{equation}
\begin{equation}\nonumber
    \tilde\kappa\ =\ \nu\ =\ \gamma = 0.
\end{equation}
At the horizon $\epsilon(r_+)=\kappa$, where $\kappa$ is the
surface gravity of the black hole.

With these preliminaries out of the way, let us now apply the general
ideas of \cite{Carlip2} to this 3-dimensional black hole.  The BTZ
space-time admits a global Killing vector
\beq
\chi=\partial_v-\Omega\partial_\phi\;,\quad\Omega=N^\phi(r_+)\;.
\label{killing}
\eeq
As in \cite{Carlip2} we now define another vector field $\rho^a$ which
is given by
\beq
\nabla_a\chi^2=-2\kappa\rho_a\;,\quad\rho^a={r\over r_+}\Big(
\partial_v+N^2\partial_r-N^\phi\partial_\phi\Big)\;.\label{rho}
\eeq
It follows that $\chi\cdot\rho=0$ and ${\cal L}_\chi\rho^a=0$
everywhere. For convenience we express both vector fields
$\chi,\rho$ in the Newman-Penrose basis up to order $(r-r_+)^2$
terms
\beq &&\chi^a=l^a+(r-r_+)(\kappa n^a+2\Omega m^a)+{\cal
O}(r-r_+)^2\;,
\\
&&\rho^a={r\over r_+}l^a-(r-r_+)\kappa n^a+{\cal O}(r-r_+)^2\;.
\label{chirho}\eeq
Clearly, at the horizon $\chi\heq\rho\heq l$. Two other useful
identities are $\nabla_a\rho_b=\nabla_b\rho_a$ and $\chi^a\nabla_a
\chi_b=\kappa\rho_b$ which follow from the definition (\ref{rho})
of $\rho$ and the fact that $\chi$ is a Killing vector.

 As in \cite{Carlip2}, the classical phase-space can be taken to be
the space of
solutions of Einstein's equations. Each space-time configuration
which is a point in the phase-space contains an inner as well as
an outer boundary. Moreover, all space-time configurations in the
neighborhood of the inner boundary are BTZ-like. For this to
achieve \cite{Carlip2} uses a set of boundary conditions which
insure that all space-times admit a Killing vector $\chi$ in a
neighborhood of the inner boundary and posses the same
`near-horizon geometry'. More precisely, it requires
\beq
\chi^a\chi^b\delta g_{ab}\heq 0\;,\quad \chi^at^b\delta g_{ab}\heq
0\label{bc}, \eeq
where $t^a$ is any space-like vector tangent to the inner boundary
($t\cdot\chi=0$). The hat over the equality sign here means that the
above equation holds on the horizon.  Clearly the vector field $\xi$
which preserves these boundary conditions (\ref{bc}) under
diffeomorphisms has to be tangent to the horizon. Keeping the same
notation as in \cite{Carlip2} let us take the vector field to be
\beq
\xi^a=T\chi^a+R\rho^a\label{vectorf} \eeq
where $R$ and $T$ are arbitrary functions. By demanding that
(\ref{vectorf}) preserves (\ref{bc}) under diffeomorphisms one puts
restrictions on $R$ and $T$. These are derived in
\cite{Carlip2} (cf. eq (4.8))
\beq
R={1\over\kappa}{\chi^2\over\rho^2}DT\;,\quad D\equiv\chi^a
\nabla_a\;.\label{rt} \eeq
The vector field, satisfying (\ref{rt}), can then be said to generate
symmetries in the precise sense of (\ref{bc}).

Let us now check the closure of the Lie-algebra of these vector
fields. It is at this point that the analysis of \cite{Carlip2}
appears to be flawed. The errors arise at three levels:
\\ a) As noted in \cite{Carlip2} the requirement that the Lie
bracket of symmetry vector fields should close imposes a new
condition
\beq {\cal
L}_\rho T \heq 0.\;\label{newcond} \eeq
In \cite{Carlip2} this condition was imposed {\em at the
horizon}. However, at the horizon $\rho^a\heq \chi^a\heq l$ and hence,
(\ref{newcond}) reads $DT\heq 0$. Then the main steps in the
calculations of \cite{Carlip2} fail to go through. In particular, the
central charge is expressed in terms of $DT$ at the horizon and
therefore vanishes identically. This in turn implies that the entropy
also vanishes identically. While the restriction on $DT$ has been
noted explicitly in \cite{Carlip2}, its (obvious) consequences on the
value of the central change and entropy are overlooked.
\\ b) Furthermore, it is {\em not} sufficient to impose
(\ref{newcond}) only {\em at} the horizon; closure will fail
unless it holds in a neighborhood.
\\ c) Later, for explicit calculations, a specific function $T$ is
chosen in \cite{Carlip2} (cf. eq. (5.6)) . Unfortunately, this
function does not satisfy the condition (\ref{newcond}) which is
required in the earlier part of the analysis in \cite{Carlip2}.

In other words, although the boundary conditions (\ref{bc}) and
(\ref{newcond}) are reasonable, the technical implementation of them,
as presented in \cite{Carlip2}, is incorrect.  In the next section we
will propose an implementation of the boundary conditions that does
not suffer from these problems.

\section{New Sets of Symmetries}

The purpose of this section is to present a systematic analysis
which is free of the technical flaws discussed above. However,
before embarking on this discussion, in section 3.1 we first
investigate a separate issue. In Carlip's analysis, the symmetry
vector fields are defined in a neighborhood of the horizon. From
general, classical considerations, one might expect that it should
be possible to focus just on the horizon structure and consider
symmetry vector fields defined intrinsically on the horizon. We
consider this possibility in Section 3.1 and show that in this
case the central charge vanishes. Thus the Carlip-type analysis
can \textit{not} be carried out with symmetries defined
intrinsically on the horizon. This result suggests that, although
the analysis appears to be classical, the origin of the central
charge ---and hence entropy--- can not be captured in a classical
analysis.  In the remainder of the section, we consider symmetry
vector fields more closely related to those of \cite{Carlip2} and
improve on that analysis.

\subsection{Geometrical symmetries}

A new framework that is naturally suited for the analysis of
symmetries on the horizon is now available -- the so-called `isolated
horizons'.  This is a notion that captures the minimum structure
intrinsic to the horizon to describe an equilibrium state of a black
hole. It allows, however, for matter and radiation in an arbitrary
neighborhood of the black hole, as long as none crosses the
horizon. As suggested in \cite{Carlip2}, this framework is well-suited
for the Carlip approach to entropy.  A comprehensive description of
the isolated horizons framework is given in \cite{letter,
afk}. Isolated horizons in 2+1 dimensions are discussed in detail in
\cite{ihtpo}.

It is natural to define symmetries as maps which preserve the basic
horizon structure, by which we mean the induced metric and a class of
null generators. More details are given in the Appendix. Here let us
just state that a vector field which preserves that structure
must be tangent to the horizon, i.e. of the form \beq \xi^a\heq
Al^a+Bm^a\label{xi} \eeq where the functions $A$ and $B$ are
restricted to be
\begin{eqnarray}
    A & = & C(v_{-})+\const. \cdot  v\\
    B & = & \const.
\end{eqnarray}
The coordinates $v$ and $v_{-}$ are defined by the relations
$n=-{\rm d}v$, $m=\frac{1}{r_{+}} \frac{\partial}{\partial \phi}$,
and $v_{\pm} = v \mp \phi/\Omega$. It is easy to see that the
algebra of these vector fields (\ref{xi}) closes. Now the boundary
conditions at the isolated horizon induce a natural symplectic
structure in the phase space, where the phase space consists of
all possible space-time configurations which admit a fixed
isolated horizon. The symplectic structure can be used to evaluate
the Poisson brackets between any two phase space functionals.

It is not difficult to check that the vector field (\ref{xi}), is
Hamiltonian. For details see the Appendix. The Poisson brackets of
the corresponding Hamiltonians close on-shell
\beq
\{H_{\xi_1},H_{\xi_2}\}\heq H_{[\xi_1,\xi_2]}\;.\label{close} \eeq
Hence, the central charge is zero.  This result is not quite
unexpected since our analysis is entirely classical and typically
the central charge arises from the failure of the classical
symmetries to be represented in the quantized theory. This shows
that, in general, for symmetries represented by smooth vector
fields on the horizon, the ideas of \cite{Strom2, Carlip2} do not
go through. If one wishes to use smooth fields ---as is most
natural at least in the classical theory--- the central charge can
arise only from quantization and the analysis would be sensitive
to the details of the quantum theory, such as the regularization
scheme used, etc. If the original intent of the ideas of
\cite{Strom2, Carlip2} is to be preserved, one must consider
symmetries represented by vector fields which do not admit smooth
limits to the horizon; in a consistent treatment, the use of
``stretched horizons'' \cite{Carlip2} is not optional but a
necessity. Perhaps this is the price one has to pay to transform
an essentially quantum analysis in the language of classical
Hamiltonian theory.

Finally, note that any reasonable local definition of a horizon should
lead to the above conclusions since we have made very weak assumptions
in this sub-section.

\subsection{Extended notion of symmetries}

Let us now return to the discussion of Section 2 and consider
symmetry vector fields defined in a neighborhood of the horizon.
Thus, we will now use the stronger set of conditions (\ref{bc})
which requires that the closure condition (\ref{newcond}) be
satisfied {\em everywhere}
\footnote{Strictly speaking, we only consider a
neighborhood of the horizon where the vector field $\chi$ is
Killing. In the BTZ example, however, it is globally Killing}.
This guarantees that the Lie-algebra of the vector fields
(\ref{vectorf}) closes \beq [\xi_{T_1},\xi_{T_2}]={\cal
L}_{\xi_{T_1}}\xi_{T_2} =\xi_{T_1DT_2-T_2DT_1}\;,\quad\xi_T=
R\rho+T\chi \eeq where $R$ is determined in terms of $T$ as in
(\ref{rt}). One is to make use of the facts that ${\cal L}_\chi
\rho={\cal L}_\rho T={\cal L}_\rho R=0$. The condition
(\ref{newcond}), however, restricts the choice of the vector
fields everywhere. To solve for the vector fields we consider a
`stretched'-horizon at $r=r_++\varepsilon$ as the inner boundary.
The solutions that are of the form \beq T_n \sim f_n(r) \exp(in
\Omega v_{+}) \label{ansatz} \eeq are especially interesting
because they furnish a Diff($S^1$), provided $f_nf_m\sim f_{n+m}$.
However, the condition (\ref{newcond}) is to be imposed carefully
because of the $(r-r_+)$ terms in the vector field $\rho$
(\ref{rho}) \beq \rho^a\nabla_a
T\sim\Big(\partial_{v_+}+N^2\partial_r\Big)T=0\;. \eeq Clearly,
the radial derivative of $T$ blows up at the horizon. With the
ansatz (\ref{ansatz}) there is a unique solution for $T_n$ in the
neighborhood of the horizon \beq T_n^\varepsilon={1\over
2\Omega}\exp\Big(-in{\Omega\over\kappa} \log(r-r_+)+in\Omega
v_+\Big)\;.\label{solved} \eeq The normalization of $T$ is so
chosen that the vector fields $\xi$ form a Diff($S^1$) algebra
\beq
[\xi_{T_n}^\varepsilon,\xi_{T_m}^\varepsilon]=i(n-m)\xi_{T_{m+n}}^
\varepsilon\label{diffs1} \eeq in the neighborhood of the horizon.

Notice that because of (\ref{solved}) the vector fields $\xi$ {\em
do not have a well defined limit at the horizon}. They are defined
only at the stretched horizon and oscillate wildly in the limit
$r\to r_+$. Also the radial derivative of $\xi$ blows up, as
expected from the condition (\ref{newcond}). So one has to take
great care in evaluating the Poisson bracket and Hamiltonians --
now {\em one cannot ignore terms which are of order} ${\cal
O}(r-r_+)$ especially in presence of radial derivatives in the
Poisson brackets. Actually more terms will contribute to the
Poisson bracket and a thorough examination of the entire
calculation is needed.

\subsection{Hamiltonian and Poisson bracket algebra}

The existence of the Hamiltonian under the boundary conditions
(\ref{bc}) is shown in \cite{Carlip2}.
 The surface Hamiltonian is (the bulk Hamiltonian is zero by constraints)
\beq
H_{\xi_n}^\varepsilon={1\over 2\pi}\oint_{S_\Delta}\epsilon_{abc}
\nabla^b\xi_n^{\varepsilon c}\;.\label{hamilt}
\eeq

The phase space, described in section 2 is associated with a
conserved symplectic current \cite{Wald}. The corresponding
symplectic structure may be used to evaluate the Poisson brackets
between any two functionals in the phase-space. On shell, the
symplectic structure can be written as the sum of boundary terms
only. However, one may choose appropriate fall-off conditions of
the fields at asymptotic infinity such that the contribution from
the outer boundary vanishes. In the present example the fields
approach `strongly' to asymptotic AdS-values. In that case given
two Hamiltonian vector fields $\xi_1$ and $\xi_2$, the Poisson
bracket between the two corresponding Hamiltonian functionals is
given solely by the terms at the inner-boundary \cite{Wald} \beq
\{H_{\xi_1},H_{\xi_2}\}=\oint_{S_\Delta}\Big(\xi_2\cdot\Theta[g,{\cal
L}_{\xi_1}g]-\xi_1\cdot\Theta[g,{\cal L}_{\xi_2}g]-\xi_2\cdot(
\xi_1\cdot L)\Big) \eeq where $2\pi\Theta_{a}[g,\delta
g]=\epsilon_{ab}[g^{bc}\nabla_c(g_{de} \delta
g^{de})-\nabla_c\delta g^{bc}]$ is the one-form symplectic
potential and $L$ is the three-form Lagrangian density. Making use
of Einstein's equations $R_{ab}=2\Lambda g_{ab}$ we can express
the Poisson bracket explicitly in terms of the vector fields \beq
\{H_{\xi_1},H_{\xi_2}\}={1\over
2\pi}\oint_{S_\Delta}\epsilon_{abc}
\Big[\xi_2^b\nabla_d(\nabla^d\xi_1^c-\nabla^c\xi_1^d)+8\Lambda
\xi_2^b\xi_1^c-(1\leftrightarrow 2)\Big]\;.\label{pb} \eeq

Our purpose is to find the terms proportional to $n^3$ in the
Poisson bracket (\ref{pb}) which give rise to a non-trivial
central extension to the Poisson bracket algebra. The Hamiltonian
(\ref{hamilt}) contains terms only linear in $n$. The central
charge can then be read off from the $n^3$ terms with appropriate
normalizations. After a long calculation we
arrive at the following 
expression \beq \lim_{\varepsilon\to
0}\Big[\{H_{\xi_n}^\varepsilon,H_{\xi_m}^
\varepsilon\}\Big]=4in^3\delta_{m+n}{a_\Delta\Omega\over 2\pi
\kappa} + \mbox{terms linear in $n$}\;.\label{n3} \eeq Notice that
although the vector fields (\ref{solved}) do not have a smooth
limit as $r\to r_+$ the Hamiltonian and the Poisson bracket have
well defined limits.
\subsection{Entropy arguments}

According to the standard normalization (up to linear order terms
in $n$) \beq \lim_{\varepsilon\to
0}\Big[\{H_{\xi_n}^\varepsilon,H_{\xi_m}^
\varepsilon\}-i(n-m)H_{\xi_{n+m}}^\varepsilon\Big]=i{c\over 12}
n^3\delta_{n+m}\label{virasoro} \eeq the central charge can be
read off from the $n^3$-term in the Poisson bracket (\ref{pb})
\beq c=24{a_\Delta\Omega\over\pi\kappa}\;.\label{cent} \eeq The
zero mode of the Hamiltonian too can be read off from
(\ref{hamilt}) and is given by \beq \lim_{\varepsilon\to
0}\Big[H_{\xi_0}^\varepsilon\Big]= {a_\Delta\kappa\over
2\pi\Omega}\;.\label{zero} \eeq Hence, by Cardy formula
\cite{Cardy}, the entropy is \beq S=2\pi\sqrt{cH_{\xi_0}\over
6}=2\sqrt 2a_\Delta \eeq which agrees with the Bekenstein-Hawking
entropy (in units $8G=\hbar=1$) up to a factor of $\sqrt 2$.

 It is worth noting here that Carlip's central extension (see
formula 5.10 of \cite{Carlip2}) and zero-th mode Hamiltonian have
the same numerical factor as ours. Nevertheless, he argues that
one should use a different, so called effective central extension,
and obtains the right numerical factor for entropy. In our case
this strategy fails since we have an extra factor of
$\Omega/\kappa$ or its inverse in front of our expressions. It
should be stressed, however, that this factor is rigidly fixed by
the requirements that the symmetry algebra closes, that it gives a
Diff$(S^1)$, and that the symmetry vector fields are periodic in
the coordinate $\phi$ with the period $2\pi$. Moreover, following
the arguments of \cite{Kutasov, Carlip1}, since within a classical
framework it is impossible to determine the value of the
Hamiltonian in the ground state of the corresponding quantum
theory, the right value of the central charge that is to be used
in the Cardy formula is not determined classically.

\section{Discussion}

The entropy calculation of \cite{Strom2} faces certain conceptual
limitations because the asymptotic symmetries may be completely
different from the horizon symmetries.  Both central charge
(\ref{cent}) and Hamiltonian (\ref{hamilt}) are quite different
from the ones found in \cite{BH1} for asymptotic infinity. Thus,
one needs an analysis restricted to the neighborhood of the
horizon. In \cite{Carlip2}, Carlip recognized this limitation and
carried out a Hamiltonian analysis using symmetries defined near
the horizon. However, as we saw in section 2, the resulting
analysis has certain technical flaws. In particular, the vector
fields which correctly incorporate the ideas laid out in the
beginning of that paper are quite different from the ones used in
the detailed analysis later on.

In section 3 we made a proposal to overcome those technical
problems and obtained a consistent formulation which implements
the previous ideas.  However, now the symmetry vector fields
(\ref{vectorf}) do not have a well-defined limit at the horizon.
Nonetheless both the Hamiltonians and their Poisson brackets are
well-defined. Furthermore, there is a central charge which,
following the reasoning of \cite{Strom2,Carlip2}, implies that the
entropy is proportional to area. While the argument has attractive
features, its significance is not entirely clear because the
vector fields generating the relevant symmetries fail to admit
well-defined limits to the horizon. Presumably, this awkward
feature is an indication that, in a fully coherent and systematic
treatment, the central charge would really be quantum mechanical
in origin and could be sensitive to certain details of
quantization, such as the regularization scheme used. Indeed, in
the detailed analysis, we had to first evaluate the Poisson
bracket and then take the limit $\lim \epsilon \rightarrow 0$ (see
expressions (\ref{zero}) and also (\ref{virasoro})), a step
typical in quantum mechanical regularization schemes. Thus, it
could well be that the awkwardness stems from the fact that,
following \cite{Strom2,Carlip2}, we have attempted to give an
essentially classical argument for a phenomenon that is inherently
quantum mechanical.

This viewpoint is supported by our analysis of section 3.1 of
symmetries corresponding to smooth vector fields. If one requires
that vector fields generating symmetries be smooth at the horizon
---a most natural condition in a fully classical setting--- we
found that the central charge would be \textit{zero}! Thus, the
fact that the vector fields do not admit a smooth limit to the
horizon is essential to the Carlip-type analysis. The fact that
one has to `push' the analysis an $\epsilon$ away from the horizon
indicates that the procedure may be a `short-cut' for a more
complete quantum mechanical regularization\footnote{Sometimes
it is argued that only a classical central charge can give rise 
to the standard expression 
$a_\Delta/4G\hbar$ of entropy and a 
central charge induced by a quantum anomaly can only give  
corrections to this expression. This, however, need
not be the case. The central charge of a truly quantum origin 
must be a dimensionless number and the only such possibility 
is $c\sim a_\Delta/G\hbar$. This would appear in the quantum
Virasoro algebra as
\beq
[\hat L_n,\hat L_m]=(n-m)\hat L_{n+m}+{c\over 12}(n^3-n)\delta_{m+n}
\eeq
where $\hat L_n$'s are now quantum operators. The eigenvalue 
of $\hat L_0$ should also be dimensionless ($\sim a_\Delta/G\hbar$).
Thus, the correct semiclassical expression of entropy can be reproduced
even when the central charge comes from the quantum theory.}.

This, however, raises some questions about the method in general:
a) How satisfactory is the classical analysis and how seriously
should one consider such vector fields? In particular, role of
such vector fields in terms of space-time geometry is far from
obvious since they are not even defined on the horizon. b) Why
should this particular algebra be the focus of attention? c) Does
the whole analysis suggest a rather transparent quantum mechanical
regularization scheme and hence, systematically constrain the
quantum theory?

The fact that our final expression of entropy differs from the
standard Hawking-Bekenstein formula by a factor of $\sqrt 2$ also
provides a test for quantum gravity theories. The value $H_{\xi_0}$
appearing in Cardy's formula is of a quantum mechanical nature. A
classical calculation may not give the right numerical value
for it. It then follows that a quantum theory of gravity will give the
correct value for the entropy provided it (a) has classical general
relativity as its low energy limit, and, (b) the expectation value
$H_{\xi_0}$ is $a_\Delta\kappa/4\pi\Omega$ (assuming $H_{\xi_{0}}$ is
well defined in quantum theory).


In spite of the limitations of this calculation, the final result
\textit{is} of considerable interest because it is not a priori
obvious that all the relevant subtleties of the full quantum
mechanical analysis can be compressed in a classical calculation
simply by stretching the physical horizon an $\epsilon$ distance
away, performing all the Poisson brackets and then taking the
limit $\epsilon \rightarrow 0$ in the \textit{final} expressions.
Note, however, that a careful treatment of technical issues that
were overlooked in \cite{Carlip2} was necessary to bring out these
features. Indeed, our analysis provides the precise sense in which
the original intention in \cite{Strom2,Carlip2} of reducing the
problem to a classical calculation is borne out in a technically
consistent fashion.

\section{Appendix}

In this appendix we define what is called a weakly isolated
horizon. It is a more general object then an isolated horizon,
however it is sufficient for our purpose of finding the symmetries
of the horizon.

Let $\Delta$ be a null hypersurface and $l$ a future pointing null
normal vector field on $\Delta$. We will denote by $[l]$ the
equivalence class of null normals which differ from $l$ only by a
multiplicative constant. Let us also introduce a one-form
$\omega_a$ defined intrinsically on $\Delta$ by:
\begin{equation}
  \nabla_{\underleftarrow{a}} l_{b} = \omega_{a} l_{b}.
\end{equation}
The arrow in the above equation denotes the pull-back to $\Delta$.

 We call a pair $(\Delta, [l])$ a weakly isolated horizon if and only
if:\\
 1. $\Delta$ is topologically $S^{1} \times \mathbf{R}$.\\
 2. The expansion $\Theta_{(l)}$ of $l$ vanishes.\\
 3. The equations of motion hold on $\Delta$. The stress-energy
 tensor $T_{ab}$ is such that $-T^{a}_{b}l^{b}$ is future directed
 and causal.\\
 4. $\mathcal{L}_{l} \omega = 0$, where $\omega$ is the one-form
 given by the equivalence class $[l]$.\\

 We will say that a vector field $\xi$ generates a symmetry of the
horizon if the flow generated by $\xi$ on the phase-space
preserves the basic structure of the horizon, namely $[l]$ and
$q$. Here, $q_{ab} \equiv g_{\underleftarrow{ab}}$. Thus we
impose,
 \begin{eqnarray}
  \mathcal{L}_{\xi} l \in [l], \\
  \mathcal{L}_{\xi} q_{ab} = 0.
 \end{eqnarray}
 It is not difficult to check that any vector field $\xi$
satisfying the above conditions can be written as
 \begin{equation}
  \xi^{a} = A l^{a} + B m^{a},
 \end{equation}
where $A=C(v_{-}) + \const. \cdot v, B=\const$, and $v_{\pm}, v$
are defined by the relations $n=-{\rm d}v$, $m=\frac{1}{r_{+}}
\frac{\partial}{\partial \phi}$, and $v_{\pm} = v \mp
\frac{\phi}{\Omega}$. As in the main text we assume that the
vector field $m^a$ is tangent to the horizon. Note that $C(v_{-})$
must be a periodic function, therefore one can perform a Fourier
analysis and find a set of modes $\xi_{n}$.

Now, using Hamiltonian considerations, one can find the
symplectic structure and Hamiltonians in the phase-space of
isolated horizons. For details see \cite{ihtpo}. The symplectic
structure on-shell is equal to
 \begin{equation}
  \Omega(\delta_{\xi},\delta) = -\frac{1}{\pi} \oint_{S_{\Delta}}
  \left[ (\xi \cdot A_{I}) \delta e^{I} + (\xi \cdot e^{I}) \delta
  A_{I} \right] + \tilde{\Omega}(\delta_{\xi}, \delta),
 \end{equation}
where $\tilde{\Omega}$ is a gauge term which is not important for
the present analysis. $A$ and $e$ are the connection one-form and
the orthonormal triad, respectively.
 Using this expression one can find the Hamiltonian corresponding to
$\xi$ as well as the Poisson bracket of two Hamiltonians. The
corresponding expressions are
 \begin{eqnarray}
  H_{\xi} & = & -\frac{1}{\pi} \oint_{S_{\Delta}} (\xi \cdot A_{I}) e^{I} + C_{\Delta}, \\
  \{H_{\xi_{1}},H_{\xi_{2}}\}& = & -\frac{1}{\pi} \oint_{S_{\Delta}} \left[ (\xi_{1}
  \cdot A_{I}) \mathcal{L}_{\xi_{2}} e^{I} + (\xi_{1} \cdot e^{I})
  \mathcal{L}_{\xi_{2}} A_{I} \right],
 \end{eqnarray}
where $C_{\Delta}$ is zero except when $\xi$ contains a constant
multiple of $l$. Then we have
$C_{\Delta}[cl]=c(M+2r_{+}\kappa+J\Omega)$.

Subsequently, one can check that for any such symmetry vector fields
\begin{equation}
\{ H_{\xi_{1}}, H_{\xi_{2}} \} \heq H_{[\xi_{1},\xi_{2}]},
\end{equation}
and therefore there is no central extension of the corresponding
algebra of conserved charges.

\vspace{-.2cm}
\subsection*{Acknowledgments}
We gratefully thank Abhay Ashtekar for discussions and various
important suggestions. Also, we would like to thank Steve Carlip
for valuable correspondence. The work of AG was supported by the
National Science Foundation grant PHY95-14240 and the Eberly
Research Funds of Penn State.

\end{document}